\documentclass[conference]{IEEEtran}
\IEEEoverridecommandlockouts
\usepackage{cite}
\usepackage{url}
\usepackage{amsmath,amssymb,amsfonts}
\usepackage{algorithmic}
\usepackage{graphicx}
\usepackage{textcomp}
\usepackage{xcolor}
\def\BibTeX{{\rm B\kern-.05em{\sc i\kern-.025em b}\kern-.08em
    T\kern-.1667em\lower.7ex\hbox{E}\kern-.125emX}}
\begin{document}
% Reduce vertical space around equations
\setlength{\abovedisplayskip}{1pt}
\setlength{\abovedisplayshortskip}{0pt}

\setlength{\belowdisplayskip}{5pt}
\setlength{\belowdisplayshortskip}{4pt}

\title{Model-Based Iterative Reconstruction with View-Dependent Detector Displacements for Cone-Beam CT
}
\author{
\IEEEauthorblockN{
Doga Topcicek\textsuperscript{1,2},
Sharmin Majumder\textsuperscript{2},
Whitney A. Poling\textsuperscript{3},
Dmitriy D. Bruder\textsuperscript{3},
Zhili Feng\textsuperscript{2},
Dali Wang\textsuperscript{2},
Xiao Wang\textsuperscript{2}
}
\IEEEauthorblockA{
\textsuperscript{1}Bredesen Center, University of Tennessee, Knoxville, TN, USA
\quad
\textsuperscript{2}Oak Ridge National Laboratory, Oak Ridge, TN, USA\\
\textsuperscript{3}General Motors Research and Development, Warren, MI, USA
}
%\thanks{Notice: This manuscript has been authored by UT-Battelle, LLC,  under contract DE-AC05-00OR22725 with the US Department of Energy (DOE).\ The US government retains and the publisher, by accepting the article for  publication, acknowledges that the US government retains a nonexclusive,  paid-up, irrevocable, worldwide license to publish or reproduce the published  form of this manuscript, or allow others to do so, for US government purposes.  DOE will provide public access to these results of federally sponsored research  in accordance with the DOE Public Access Plan(\url{https://www.energy.gov/doe-public-access-plan}).}
}
\maketitle

\vspace{-5pt}
\begin{abstract}
Detector shifting is used in industrial cone-beam computed tomography to reduce ring artifacts, but the resulting detector motion can vary nonuniformly across projection views and produce detector positions that cannot be represented by fixed offsets. Although analytical reconstruction can accommodate this motion, approaches that use projection interpolation or rebinning may limit spatial resolution and introduce artifacts. MBIR provides an alternative by representing the acquisition geometry explicitly within the reconstruction model. Existing separable distance-driven MBIR formulations assume fixed detector geometry across projection views and therefore cannot directly represent nonuniform motion in both detector dimensions, as this motion changes the voxel-to-detector mapping. We extend the separable distance-driven MBIR model to incorporate recorded horizontal and vertical detector displacements into the distance-driven overlap computation while leaving measured projections unchanged. Forward-projection validation achieved an NRMSE of 0.0338 across 1080 views. The proposed model achieved a 3D SSIM of 0.9404 and NRMSE of 0.0406, compared with 0.5396 and 0.1740 for fixed-detector MBIR. A second industrial dataset showed reduced structured artifacts compared with FDK.
\end{abstract}

\vspace{-5pt}
\section{\textbf{Introduction}}
\label{sec:introduction}
Cone-beam CT (CBCT) reconstructs a three-dimensional volume from X-ray
projections acquired with a two-dimensional detector
\cite{scarfe2008cone}. Accurate reconstruction requires the acquisition geometry to be represented precisely. Model-based iterative reconstruction (MBIR), unlike analytical methods such as the
Feldkamp--Davis--Kress (FDK) algorithm \cite{feldkamp1984practical},
incorporates acquisition geometry along with physical and statistical models into
reconstruction \cite{thibault2007three,wang2017massively}. Because the MBIR forward model uses the scanner geometry to map the image volume to detector measurements, errors in the assumed geometry can create a mismatch between the predicted and measured projections \cite{thibault2007three,balke2018separable}.

Detector shifting has been used to reduce ring artifacts, which can arise
from miscalibrated or defective detector elements \cite{boas2012ct}. In
industrial CT inspection, such artifacts can reduce the visibility of
internal structures and make defects more difficult to identify \cite{hang2025ring}. This detector motion causes object features
to be sampled by different detector elements across views rather than at the
same detector locations \cite{zhu2013micro,liu2023detector}. In the acquisitions considered here, the detector undergoes nonuniform
horizontal and vertical movement across views. A pair of fixed detector offsets cannot represent these changing detector positions, and ignoring the motion introduces a mismatch between the acquisition and reconstruction models.

In analytical reconstruction methods such as FDK, detector displacement can be accommodated through coordinate correction, projection interpolation, or rebinning, although such processing methods can limit spatial resolution and contribute to reconstruction artifacts \cite{wang2021physics}. In MBIR, however, the changing detector geometry must be represented directly within the forward model. Because each detector position
changes how the projected footprint of each voxel overlaps the detector elements, the system matrix must account for the measured detector position at each view. This makes detector displacement more complex to represent in MBIR than in FDK, but explicitly modeling the acquisition geometry can improve spatial fidelity and reduce ring artifacts.
Existing separable distance-driven MBIR formulations based on fixed detector geometry cannot directly represent measured
nonuniform detector motion in both detector dimensions across projection views. We therefore extend the separable distance-driven projector to incorporate the measured nonuniform horizontal and vertical detector displacements into the per-view
detector-element coordinates, while leaving the measured projection data unchanged.

\section{\textbf{Background and Related Work}}
\label{sec:background}
\subsection{\textbf{Detector Motion and Artifact-Correction Approaches}}
Zhu et al. addressed artifacts caused by detector defects and nonuniform
detector response by combining intentional detector shifting with data
inpainting to correct projection data \cite{zhu2013micro}. Liu et al. introduced
random horizontal detector shifts during acquisition and used the resulting
reconstructed CT images as input to a U-Net for ring-artifact and noise
suppression \cite{liu2023detector}. These approaches modify projection data or perform post-reconstruction suppression, whereas our method incorporates recorded detector displacements directly into the distance-driven MBIR geometry.

\subsection{\textbf{Geometric Calibration and Correction}}
Other methods address differences between scanner and reconstruction geometry. Ferrucci et al. showed that detector tilt, slant, and skew can introduce
reconstruction errors and corrected the resulting radiographic distortions
by resampling simulated radiographs using distortion maps and linear
interpolation \cite{ferrucci2016evaluating}. Cho et al. used a
steel-ball calibration phantom to estimate source and detector positions, detector orientation, and other geometric parameters \cite{cho2005accurate}. Nguyen used
a metal-ball calibration phantom to estimate a projection matrix
independently at each projection view, allowing the acquisition geometry to
be represented for each view \cite{nguyen2016view}. Unlike these correction and calibration approaches, the detector displacements considered here are already recorded during acquisition and must be incorporated into the reconstruction model.

\subsection{\textbf{Acquisition Geometry in Reconstruction Models}}
Balke et al. described a cone-beam MBIR system model using fixed detector
geometry parameters across projection views \cite{balke2018separable}. Tseng et al. investigated cone-beam CT reconstruction using fixed lateral detector offsets \cite{tseng2022cone}. These formulations
represent nominal or fixed detector positions rather than detector positions
that vary across projection views. Wang et al. modeled flying-focal-spot CT by incorporating view-dependent
focal-spot locations directly into the reconstruction geometry
\cite{wang2021physics}. Their focal-spot motion follows a defined geometric
relationship with scanner rotation, whereas the present work incorporates
recorded nonuniform detector motion in both detector dimensions. These formulations address fixed detector offsets or prescribed source motion, but do not model recorded detector motion across views in a separable cone-beam distance-driven MBIR projector.

\subsection{\textbf{Distance-Driven MBIR Formulation}}
\label{sec:distance_driven}
\vspace{-10pt}
\begin{figure}[!htbp]
    \centering
    \includegraphics[
        width=\columnwidth,
        trim={7.25cm 5.5cm 8.95cm 3.8cm},
        clip
    ]{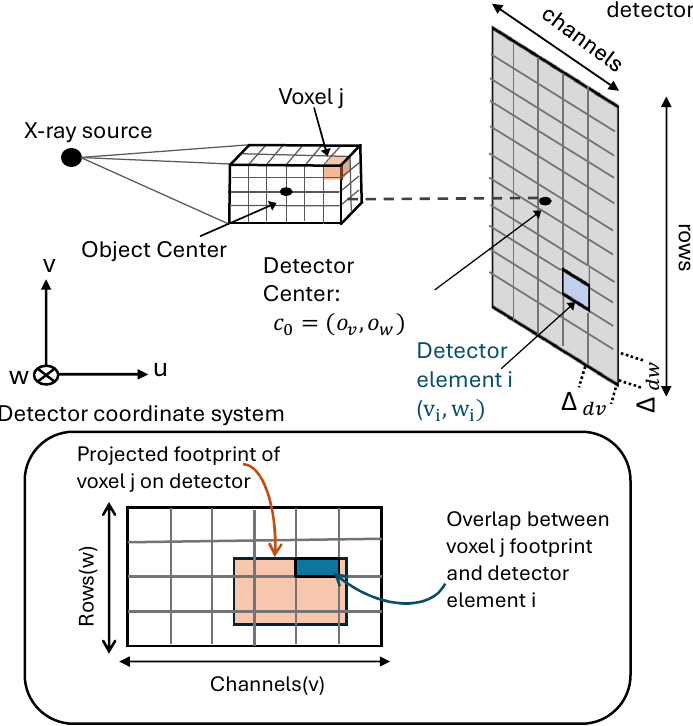}
    \caption{Voxel-to-detector overlap in the distance-driven projector.}
\label{fig:distance_driven_overlap}
\end{figure}
As shown in Fig.~\ref{fig:distance_driven_overlap}, the scanner coordinate
system is defined by the $u$-, $v$-, and $w$-directions. The $u$-direction
extends from the X-ray source toward the detector, with $v$ aligned with
the detector channels and $w$ aligned with the detector rows. The nominal
detector center in the $v$-$w$ plane is $\mathbf{c}_0=(o_v,o_w)$, 
where $o_v$ and $o_w$ are the global detector offsets in the channel and row
directions, respectively. In the fixed-detector baseline, these offsets
remain constant across all projection views.

The measured projection data are modeled as
\begin{equation}
    \mathbf{y}
    =
    \mathbf{A}\mathbf{x}
    +
    \boldsymbol{\epsilon},
    \label{eq:forward_model}
\end{equation}
and the reconstructed image is estimated by minimizing a weighted
data-fidelity term together with an image regularization term,
\begin{equation}
\small
\widehat{\mathbf{x}}
=
\arg\min_{\mathbf{x}\geq0}
\left\{
\frac{1}{2}
(\mathbf{y}-\mathbf{A}\mathbf{x})^T
\mathbf{W}
(\mathbf{y}-\mathbf{A}\mathbf{x})
+
R(\mathbf{x})
\right\}.
\label{eq:mbir_objective}
\end{equation}
Here, $\mathbf{x}$ is the image volume, $\mathbf{y}$ is the measured
projection data, $\boldsymbol{\epsilon}$ represents measurement and modeling
error, $\mathbf{W}$ is the measurement-weight matrix, and $R(\mathbf{x})$
is the image regularization term. The system matrix $\mathbf{A}$ describes
the mapping from the image volume to the projection measurements according
to the acquisition geometry \cite{wang2026efficient}.

In the distance-driven projector, each voxel is projected onto the detector
to form a footprint, and its system-matrix contribution is determined from
the overlap between the projected footprint and the detector elements
\cite{de2004distance}. Let $i=(i_\beta,i_v,i_w)$ denote a projection
measurement indexed by projection view, detector channel, and detector row,
and let $j=(j_x,j_y,j_z)$ denote an image voxel. In the separable cone-beam
formulation, $B_{i,j}$ and $C_{i,j}$ denote the channel- and row-direction
contributions, respectively, and each system-matrix coefficient is factored as
\cite{balke2018separable},
\begin{equation}
A_{i,j}=B_{i,j}C_{i,j}.
\label{eq:separable_system_matrix}
\end{equation}
\section{\textbf{Distance-Driven MBIR with View-Dependent Detector Displacements}}

The proposed model retains the separable distance-driven structure while representing the recorded detector displacements through a view-dependent system matrix:
\begin{equation}
\mathbf{y}(\Delta v,\Delta w)
=
\mathbf{A}(\Delta v,\Delta w)\mathbf{x}
+
\boldsymbol{\epsilon}.
\label{eq:view_dependent_forward_model}
\end{equation}
Here, $\Delta v$ and $\Delta w$ denote the recorded
channel- and row-direction detector displacements across projection views.
The notation $\mathbf{y}(\Delta v,\Delta w)$
denotes projection data acquired under the recorded detector geometry; the
measured projection values remain unchanged, while the system matrix
$\mathbf{A}(\Delta v,\Delta w)$ accounts for the
displacements through the voxel-to-detector geometry.
\begin{figure}[!h]
    \centering
    \includegraphics[
    width=\columnwidth,
        trim={4.7cm 5.75cm 9.5cm 6.4cm},
        clip
    ]{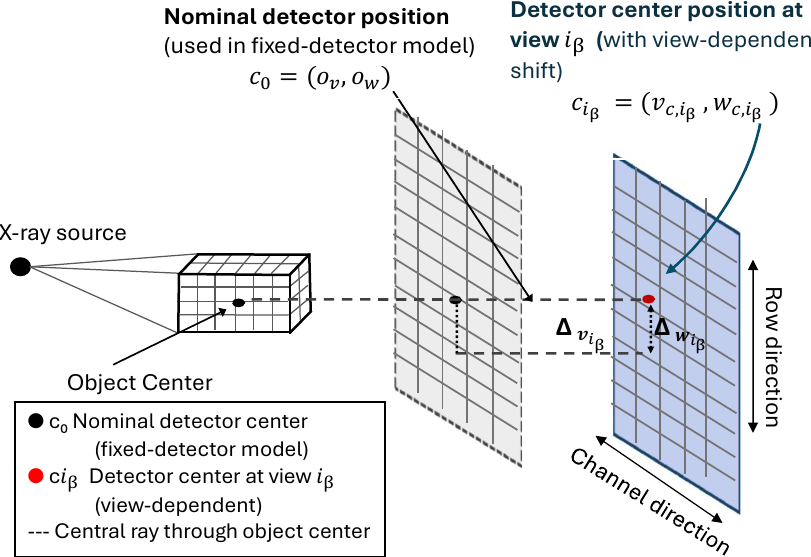}
   \caption{Recorded detector displacements $\Delta v_{i_\beta}$ and
$\Delta w_{i_\beta}$ modify the nominal detector center $\mathbf{c}_0$
to obtain the detector center $\mathbf{c}_{i_\beta}$ at projection view
$i_\beta$.}
\label{fig:view_dependent_geometry}
\end{figure}
At projection view $i_\beta$, $\Delta v_{i_\beta}$ and
$\Delta w_{i_\beta}$ denote the channel- and row-direction detector
displacements, respectively, obtained by mapping the recorded left/right
and up/down motion to the $v$- and $w$-directions. As shown in Fig.~\ref{fig:view_dependent_geometry}, these displacements modify the
nominal detector center. The resulting detector center
$\mathbf{c}_{i_\beta}=(v_{c,i_\beta},w_{c,i_\beta})$ is given by
\begin{equation}
v_{c,i_\beta}=o_v+\Delta v_{i_\beta}, \qquad
w_{c,i_\beta}=o_w+\Delta w_{i_\beta}.
\label{eq:view_dependent_detector_center}
\end{equation}
Here, $o_v$ and $o_w$ are global offsets and remain fixed across projection views.
For detector spacings $\Delta_{dv}$ and $\Delta_{dw}$, the center
coordinates of detector element $(i_v,i_w)$ at projection view $i_\beta$
are
\vspace{-10pt}
\begin{equation}
\begin{aligned}
v_{i_v,i_\beta}
&=
v_{c,i_\beta}
+
\left(
i_v-\frac{N_{dv}-1}{2}
\right)\Delta_{dv}, \\
w_{i_w,i_\beta}
&=
w_{c,i_\beta}
+
\left(
i_w-\frac{N_{dw}-1}{2}
\right)\Delta_{dw},
\end{aligned}
\label{eq:view_dependent_detector_grid}
\end{equation}
where $N_{dv}$ and $N_{dw}$ are the numbers of detector channels and rows, respectively. Thus, the recorded displacements produce per-view detector-element coordinates in both detector dimensions, which determine the channel- and row-direction overlap contributions.
For the channel direction, let
$\bar{v}_{j}^{(i_\beta)}$ denote the detector-channel coordinate of the
projected center of voxel $j$ at projection view $i_\beta$, and let
$W_{pv,j}^{(i_\beta)}$ denote its projected footprint width. The recorded
channel-direction detector displacement changes the detector-channel coordinate
$v_{i_v,i_\beta}$ used in the channel-direction overlap calculation.
The distance from the center of detector channel $i_v$ to the projected voxel center is \begin{equation}
\delta_{v,i,j}
=
\left|
v_{i_v,i_\beta}
-
\bar{v}_{j}^{(i_\beta)}
\right|.
\label{eq:channel_center_distance}
\end{equation}
The corresponding channel-direction overlap length is
\begin{equation}
\resizebox{0.88\columnwidth}{!}{$\displaystyle
L_{v,i,j}
=
\max\left\{
\frac{W_{pv,j}^{(i_\beta)}+\Delta_{dv}}{2}
-
\max\left(
\frac{\left|W_{pv,j}^{(i_\beta)}-\Delta_{dv}\right|}{2},
\delta_{v,i,j}
\right),
0
\right\}
$}
\label{eq:channel_overlap}
\end{equation}
and the channel-direction contribution is
\begin{equation}
B_{i,j}
=
\frac{\Delta_{xy}}
{\cos\!\left(\alpha_{xy,j}^{(i_\beta)}\right)\Delta_{dv}}
L_{v,i,j},
\label{eq:channel_factor}
\end{equation} where $\Delta_{xy}$ is the in-plane voxel spacing and
$\alpha_{xy,j}^{(i_\beta)}$ is the conventional distance-driven
in-plane footprint angle.

The channel-direction displacement therefore propagates through
$v_{i_v,i_\beta}$ and $\delta_{v,i,j}$ to modify the channel overlap
$L_{v,i,j}$ and system-matrix contribution $B_{i,j}$.

The row-direction contribution uses an intermediate $u$-coordinate in the
separable distance-driven formulation. Let $u_{j_u}$ denote the intermediate
$u$-coordinate corresponding to the projected in-plane voxel position. With
detector-plane coordinate $u_{d0}$ and source coordinate $u_s$, the
corresponding magnification is \begin{equation}
M_{j_u}
=
\frac{u_{d0}-u_s}
     {u_{j_u}-u_s}.
\label{eq:row_magnification}
\end{equation}
For voxel row coordinate $w_j$ and voxel spacing $\Delta_z$, the projected
voxel-center coordinate and projected row-direction footprint width are
\begin{equation}
\begin{aligned}
\bar{w}_{j}^{(j_u)} &= M_{j_u}w_j, \\
W_{pw,j_u} &= M_{j_u}\Delta_z.
\end{aligned}
\label{eq:row_projected_geometry}
\end{equation}
The recorded row-direction detector displacement changes the detector-row
coordinate $w_{i_w,i_\beta}$ used in the row-direction overlap calculation.
The distance between the center of detector row $i_w$ and the projected
voxel center is
\begin{equation}
\delta_{w,i,j}
=
\left|
w_{i_w,i_\beta}
-
\bar{w}_{j}^{(j_u)}
\right|.
\label{eq:row_center_distance}
\end{equation}
The corresponding row-direction overlap length is
\begin{equation}
\resizebox{0.88\columnwidth}{!}{$\displaystyle
L_{w,i,j}
=
\max\left\{
\frac{W_{pw,j_u}+\Delta_{dw}}{2}
-
\max\left(
\frac{\left|W_{pw,j_u}-\Delta_{dw}\right|}{2},
\delta_{w,i,j}
\right),
0
\right\}
$}
\label{eq:row_overlap}
\end{equation}
and the corresponding row-direction contribution is
\begin{equation}
C_{i,j}
=
\frac{L_{w,i,j}}{\Delta_{dw}}
\sqrt{
1+
\frac{w_j^2}
     {(u_{j_u}-u_s)^2}
}.
\label{eq:row_factor}
\end{equation}
The recorded row-direction displacement $\Delta w_{i_\beta}$ changes
$w_{i_w,i_\beta}$ and therefore the row overlap $L_{w,i,j}$ and contribution
$C_{i,j}$. Thus, the recorded detector displacements modify both overlap
contributions through the per-view detector coordinates while leaving the
measured projection data unchanged.

\section{\textbf{Experimental Setup}}
\label{sec:experiments}
The proposed MBIR model was evaluated on two industrial CBCT
datasets provided by General Motors: an aluminum foil-tab weld and a
3D-printed fixture. Both datasets included measured projection data, an FDK reconstruction, acquisition geometry parameters, and per-view detector-displacement data, and were acquired using a North Star Imaging industrial CBCT system with an XRayWorX P18-542 X-ray source. For each acquisition, a detector-position file provides the left/right and up/down displacements recorded in millimeters at each view during acquisition. The vendor-provided FDK reconstruction was used for comparison because it represents the established commercial reconstruction for this system and accounts for the recorded per-view detector positions. Key acquisition parameters are listed in Table~\ref{tab:acq_params}.
\vspace{-12pt}
\begin{table}[!htbp]
\centering
\caption{Acquisition parameters for the weld and fixture datasets.}
\label{tab:acq_params}
\begin{tabular}{lcc}
\hline
\textbf{Parameter} & \textbf{Weld} & \textbf{Fixture} \\
\hline
Projection views & 1080 & 2520 \\
Detector pixel pitch [mm] & 0.127 & 0.200 \\
Source-to-detector distance [mm] & 350 & 623.173 \\
Source-to-rotation-axis distance [mm] & 54.62 & 49.878 \\
Magnification & 6.408 & 12.49 \\
Detector dimensions [rows $\times$ channels] &
$1880\times1496$ & $2008\times2008$ \\
\hline
\end{tabular}
\end{table}
\vspace{-15pt}
\section{\textbf{Results}}
\label{sec:results}
\subsection{\textbf{Geometry and Forward-Projection Validation}}
\begin{figure}[!htbp]
    \centering
    \includegraphics[
        width=\columnwidth,
        trim={0.4cm 0.5cm .5cm 0.5cm},
        ]{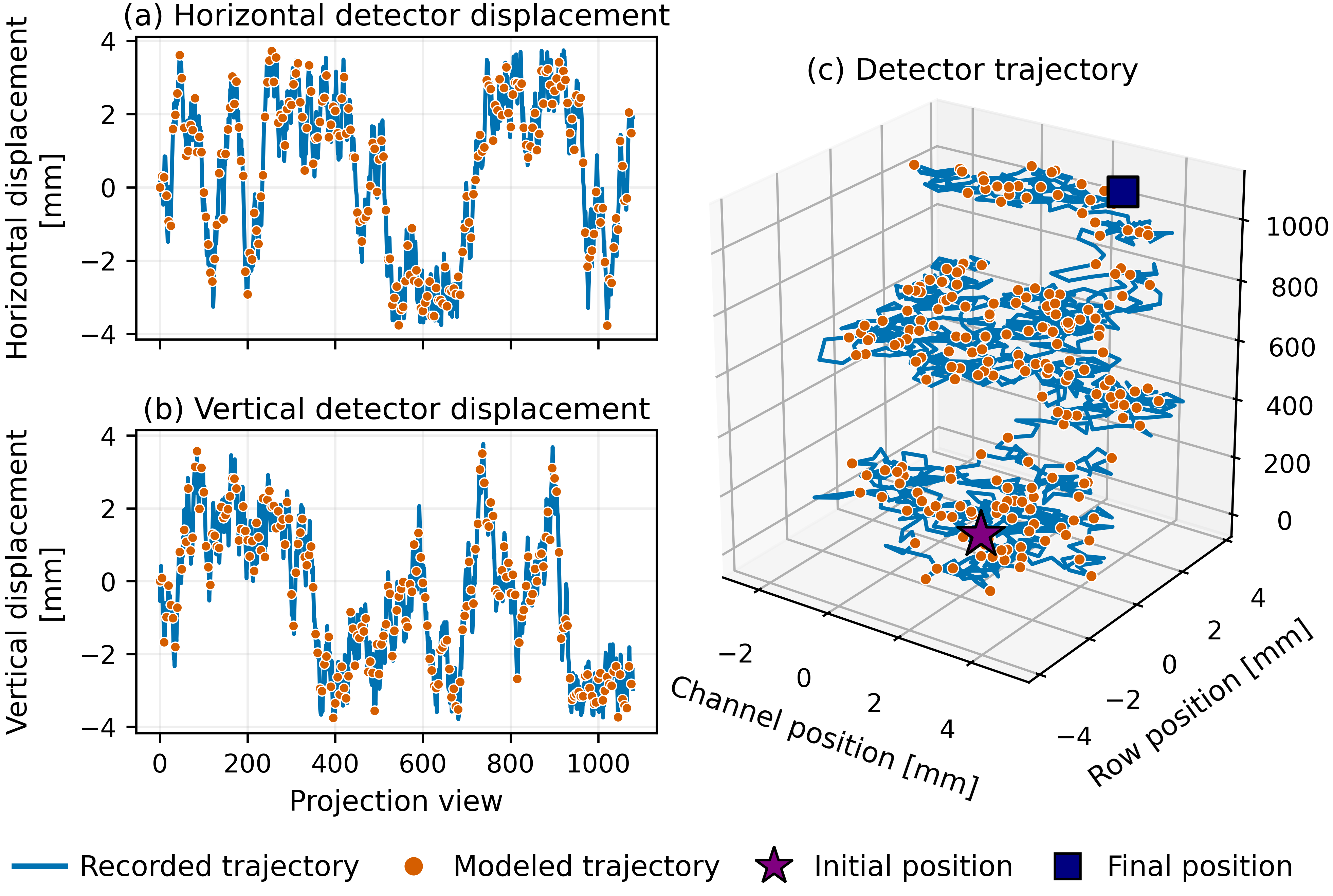}
\caption{
Recorded and modeled detector motion across projection views: (a) horizontal displacement, (b) vertical displacement, and (c) detector trajectory.}
\label{fig:detector_motion}
\end{figure}

The recorded horizontal and vertical displacements were compared with the modeled displacements to verify that the proposed geometry reproduces the recorded detector motion. Figure~\ref{fig:detector_motion} shows the recorded and modeled detector motion across the 1080 views of the weld dataset. The modeled horizontal and vertical displacements overlap the recorded values, and the resulting trajectory preserves the motion after the global detector offsets are included. This confirms that the recorded detector displacements are correctly incorporated into the proposed geometry.

To validate the geometry against the acquisition, the vendor-provided FDK reconstruction, which accounts for the recorded per-view detector positions, was forward projected using the proposed geometry to generate an emulated sinogram for comparison with the measured sinogram. For the weld dataset, NRMSE was computed over the comparison ROI across all views as the RMSE between the two sinograms divided by the intensity range of the measured sinogram, expressing the error relative to the measured-data range. The resulting NRMSE of 0.0338 indicates that the proposed system matrix accurately represents the recorded view-dependent acquisition geometry.
\vspace{-5pt}

\vspace{-15pt}
\subsection{\textbf{Reconstruction Comparison}}
\label{sec:reconstruction_quality}
For the weld dataset, detector-motion modeling was evaluated using two MBIR configurations. The fixed-detector baseline ignored recorded per-view displacements and used the same global detector offsets across all views. The proposed MBIR incorporated the recorded channel- and row-direction detector displacements at each view while leaving the measured projection data unchanged. All other reconstruction settings were unchanged. Quantitative comparison was performed over the same 3D weld ROI relative to the vendor-provided FDK reconstruction, with results summarized in Table~\ref{tab:recon_metrics}.
\vspace{-12pt}
\begin{table}[!htbp]
\centering
\caption{Quantitative reconstruction comparison for the weld dataset.}
\label{tab:recon_metrics}
\begin{tabular}{lcc}
\hline
\textbf{Method} & \textbf{3D SSIM $\uparrow$} & \textbf{3D NRMSE $\downarrow$} \\
\hline
Fixed-detector MBIR & 0.5396 & 0.1740 \\
Proposed MBIR & \textbf{0.9404} & \textbf{0.0406} \\
\hline
\end{tabular}
\end{table}
\vspace{-8pt}

\begin{figure}[!htbp]
\centering
\includegraphics[
width=\columnwidth,
trim={0.0cm 0.1cm 0.5cm 0.58cm},
clip
]{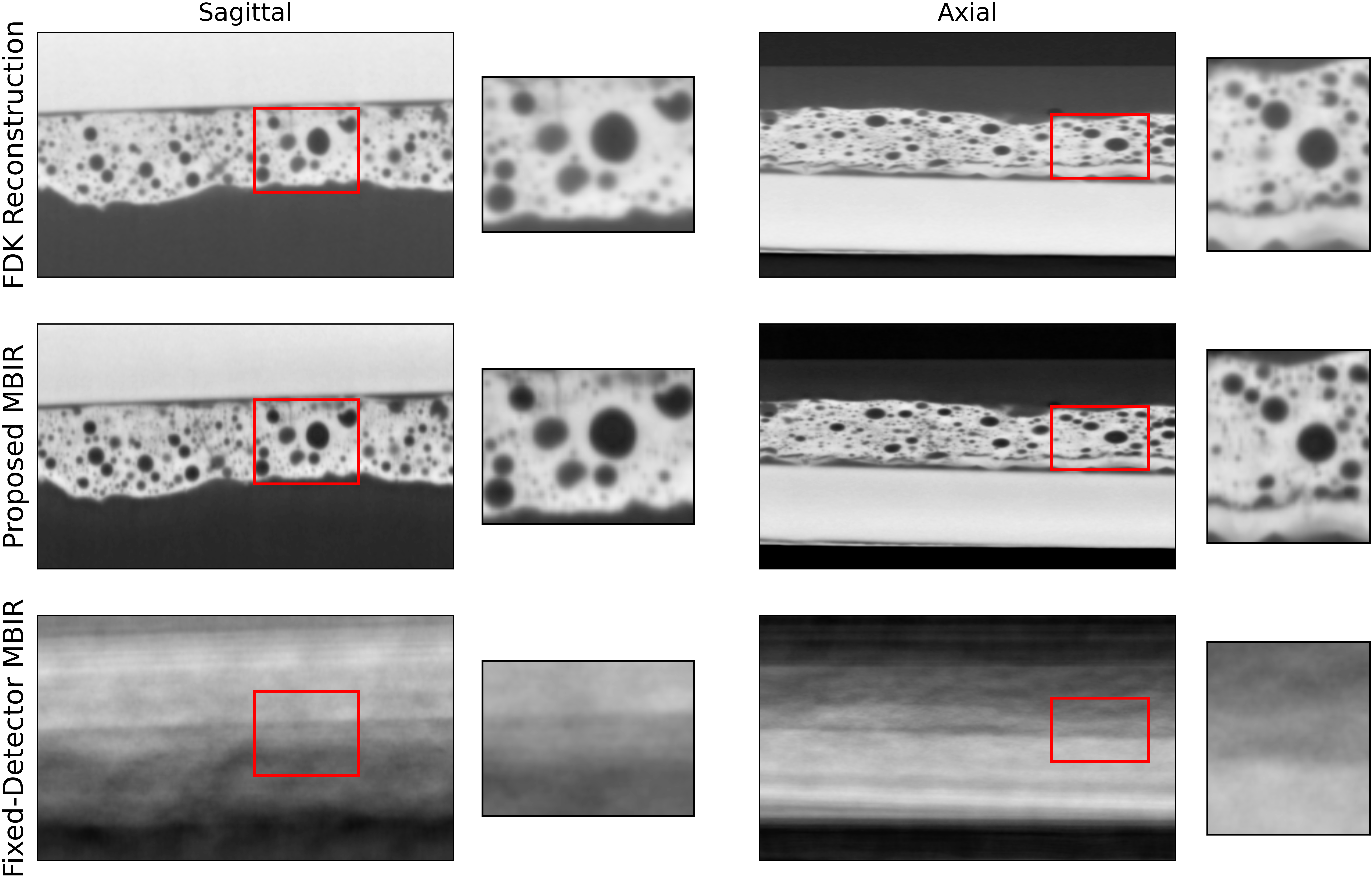}
\vspace{-20pt}
\caption{Weld reconstruction comparison: sagittal (left) and axial (right). Rows show the vendor-provided FDK reconstruction, proposed MBIR with recorded per-view detector displacements, and fixed-detector MBIR with global offsets only, from top to bottom. Red boxes mark the corresponding close-up regions.}
\label{fig:reconstruction_comparison}
\end{figure}

\begin{figure}[!htbp]
\centering
\includegraphics[
width=0.80\columnwidth
]{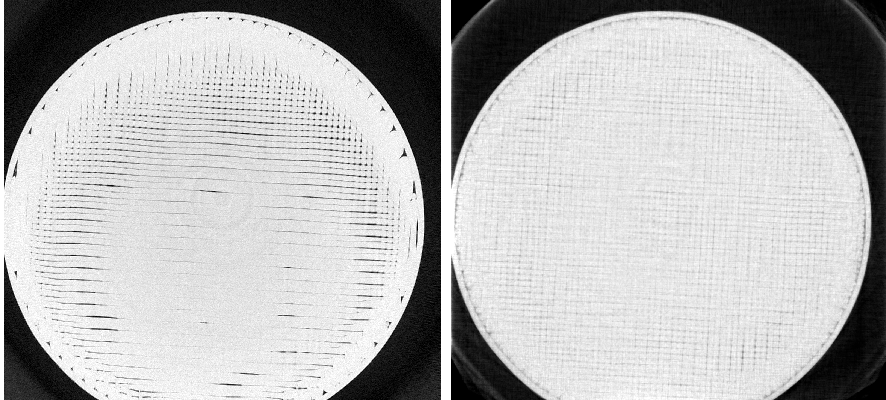}
\caption{Fixture reconstruction comparison: FDK reconstruction (left) and proposed MBIR (right).}
\label{fig:fixture_comparison}
\end{figure}

As shown in Fig.~\ref{fig:reconstruction_comparison}, ignoring the recorded per-view detector displacements produces substantial blurring and loss of local structural detail. Incorporating the detector motion preserves the pore structure and material boundaries visible in the vendor-provided FDK reconstruction, while the close-up views show improved local feature definition compared with the fixed-detector reconstruction. The proposed MBIR also shows sharper local features than the FDK reconstruction. Together with the forward-projection validation, these results demonstrate the importance of representing the recorded detector motion directly within the MBIR reconstruction geometry.
\vspace{-5pt}
Results on the second dataset, the 3D-printed fixture, show that the benefit of modeling detector motion extends beyond the weld dataset. As shown in Fig.~\ref{fig:fixture_comparison}, the proposed MBIR reconstruction reduces the structured artifacts observed in the FDK reconstruction, including the central ring-like artifact, producing a more uniform fixture interior.
\vspace{-8pt}
\vspace{-14pt}
\section{\textbf{Discussion and Conclusion}}
\label{sec:conclusion}
This work introduced a view-dependent distance-driven MBIR system model that
incorporates recorded nonuniform detector motion in both detector dimensions
directly into the voxel-to-detector overlap computation while leaving the
measured projection data unchanged. Forward-projection validation showed that
the proposed geometry closely represents the detector-shifted acquisition,
with an NRMSE of 0.0338 across 1080 measured views. Compared with
fixed-detector MBIR, incorporating the recorded per-view detector positions
increased 3D SSIM from 0.5396 to 0.9404 and reduced 3D NRMSE from 0.1740 to
0.0406. For the fixture dataset, the proposed MBIR reconstruction showed
visibly reduced structured artifacts compared with the FDK reconstruction,
including the central ring-like artifact. Together, these results demonstrate
the benefit of incorporating recorded detector motion directly into the MBIR
reconstruction geometry.

\bibliographystyle{IEEEtran}
\bibliography{references}

\end{document}